\documentclass[fleqn,usenatbib]{mnras}

\usepackage{newtxtext,newtxmath}

\usepackage[T1]{fontenc}
\newcommand{\Fig}{Figure~\ref}

\DeclareRobustCommand{\VAN}[3]{#2}
\let\VANthebibliography\thebibliography
\def\thebibliography{\DeclareRobustCommand{\VAN}[3]{##3}\VANthebibliography}

\usepackage{graphicx}	
\usepackage{amsmath}	

\title[Kinematic Structures of HI Filaments]{The Kinematic Structure of Magnetically Aligned HI Filaments}

\author[D. A. Kim et al.]{
Doyeon A. Kim,$^{1}$\thanks{E-mail: d.kim3@columbia.edu}
S. E. Clark,$^{2,3}$
M. E. Putman,$^{1}$
and Larry Li$^{1}$
\\
$^{1}$Department of Astronomy, Columbia University, New York, NY 10027, USA\\
$^{2}$Department of Physics, Stanford University, Stanford, California 94305, USA\\
$^{3}$Kavli Institute for Particle Astrophysics \& Cosmology, P. O. Box 2450, Stanford University, Stanford, CA 94305, USA 
}

\date{Accepted XXX. Received YYY; in original form ZZZ}

\begin{document}
\label{firstpage}
\pagerange{\pageref{firstpage}--\pageref{lastpage}}
\maketitle

\begin{abstract}
 We characterize the kinematic and magnetic properties of HI filaments located in a high Galactic latitude region ($165^\circ < \alpha < 195^\circ$ and $12^\circ < \delta < 24^\circ$). We extract three-dimensional filamentary structures using \texttt{fil3d} from the Galactic Arecibo L-Band Feed Array HI (GALFA-HI) survey 21-cm emission data. Our algorithm identifies coherent emission structures in neighboring velocity channels.
Based on the mean velocity, we identify a population of local and intermediate velocity cloud (IVC) filaments. We find the orientations of the local (but not the IVC) HI filaments are aligned with the magnetic field orientations inferred from Planck 353 GHz polarized dust emission. 
We analyze position-velocity diagrams of the velocity-coherent filaments, and find that only 15 percent of filaments demonstrate significant major-axis velocity gradients with a median magnitude of 0.5 km s$^{-1}$ pc$^{-1}$, assuming a fiducial filament distance of 100 pc. We conclude that the typical diffuse HI filament does not exhibit a simple velocity gradient. The reported filament properties constrain future theoretical models of filament formation.  

\end{abstract}



\section{Introduction}
Filamentary structures thread the Milky Way on almost every length scale \citep{hershel_filament2010, hacar_filform, zucker_skeletonmw, kalberla_filaments, clark_hensley}. These linear structures are observed in various molecular clouds and their ubiquity may be linked to the physics of star formation \citep{orion_filament, hi_gal, Arzoumanian2011, palmeirim_striation, hacar_sf}. Recent observations confirm that a similar intricate network of filaments exists in HI clouds in a range of Galactic environments \citep{MC_hicloud, clark_rht, HI4PI, galfa, soler_thor}.

\begin{figure*}
    \centering
    \includegraphics[width=\linewidth]{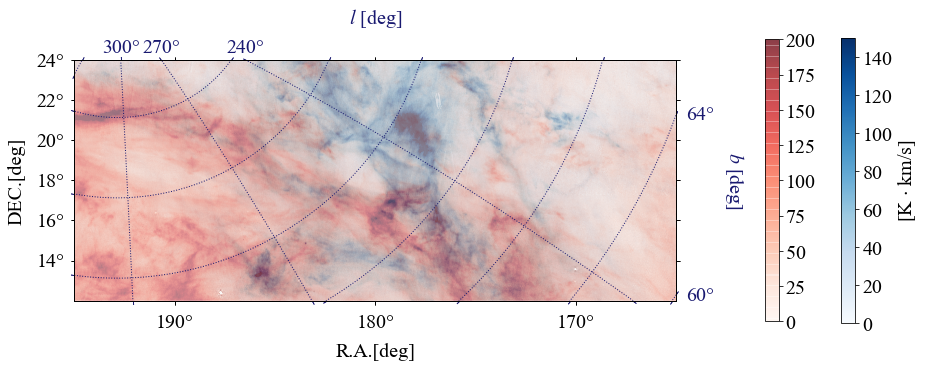}
    \caption{An overlay of integrated intensity (moment 0) maps evaluated at different velocity ranges gridded on both ICRS and galactic projections. The blue plot shows the moment 0 evaluated in the velocity range [-50, -20] km/s and the red plot shows the moment 0 evaluated from [-20, 20] km/s. }
    \label{fig:moment0}
\end{figure*}

Despite their prevalence, the detailed physics that shapes filamentary structures is not well-understood. There is some evidence that the magnetic field plays an important role, as non-self-gravitating filaments are observed to be aligned with the magnetic field in both dust and atomic gas \citep{MD_Hershel, clark_rht, panopoulou_striation}. 
Diffuse HI filaments are particularly well-aligned with the ambient magnetic field traced by both starlight polarization \citep{clark_rht} and polarized dust emission  \citep{clark-pol, kalberla_filaments}. Similar behavior is observed in low column density dusty filaments \citep{alignment_planckxxxii}; at a higher column density, the relative alignment between the magnetic field and filament long axes trends toward perpendicular \citep{soler_bfield_orientation, alignment_planckxxxii, g47sophia}.

Various physical mechanisms for filament formation have been proposed. Filaments of cold gas in the warm ISM, for instance, are proposed as a product of thermal instability and turbulent compression and shear \citep{filform_heitsch, Inoue_Inutsuka}. The joint influences of turbulence and magnetic fields can form thin, elongated density structures \citep{smith_filform, gazol_hisim}. Furthermore, stretching induced by turbulence alone can form filaments that are confined by the Lorentz force \citep{Hennebelle_hifil, magnetic_mc, Seifried_orientation_mc}. Because a filamentary geometry can result from a variety of physical scenarios, the fact of filamentarity does not on its own specify the filament formation mechanism; there may be multiple mechanisms operating across interstellar environments \citep[e.g.,][]{hacar_sf}. In addition to its morphology, the kinematic structure of filaments can help to constrain the physics of their formation.

Although a comprehensive analysis of diffuse HI filament kinematics has not yet been disclosed, some visually-inspected tentative velocity gradients along HI filaments are reported \citep{kalberla_filaments}. The kinematic structure of filaments has been analyzed more thoroughly in molecular environments. 
Clear velocity gradients are sometimes identified in molecular filaments, either running length-wise along the filament long axis \citep{Dobashi_COgrad}, or along the orthogonal axis \citep{fernandez_vgrad}. Non-self-gravitating filaments (``striations") in the vicinity of a dense filament of the Taurus molecular cloud display a large-scale velocity gradient suggestive of accretion along the striations \citep{goldsmith_striation, fil_balign}. 
The kinematic description of some filaments depends on both the spatial scales and conditions. For instance, \cite{hacar_filform} suggests that the Taurus B213 filament is actually composed of many distinct velocity structures, while a level of turbulence and type of gas tracers influence the alignment level of the filament \citep{heyer_taurus}. To further search for clues on the formation and evolution of filamentary structures, in this paper we examine the kinematics of HI filaments.

To access the kinematics of HI filaments, we utilize three-dimensional data from the Galactic Arecibo L-Band Feed Array HI (GALFA-HI) survey \citep{galfa}. As \cite{clark_rht} demonstrated, angular resolution and sensitivity are critical for identifying and characterizing slender HI filaments. We use GALFA-HI's highest available spatial and spectral resolution (4$'$ and $0.184$ $\rm{km/s}$ respectively) to study coherent structures in position-position-velocity space, i.e. the kinetic structure of 3D filaments \citep{beaumont_ppv, clark_intensity}. 


In this paper, we investigate the kinematic properties and magnetic field orientation of filamentary HI structures at high Galactic latitudes. The paper is organized as follows. In Sections \ref{sec:data} and \ref{sec:def}, we introduce the data and outline the steps to extract the 3D HI filaments. Section \ref{bigsec:methods} discusses our methods to examine the magnetic field orientation and kinematic properties of individual filaments. In Section \ref{sec:results}, we present our results. 
Finally, we discuss the possible implications of our results in Section \ref{sec:discuss} and conclude in Section \ref{sec:conclusion}.

\begin{figure*}
\centering
\includegraphics[width=\textwidth]{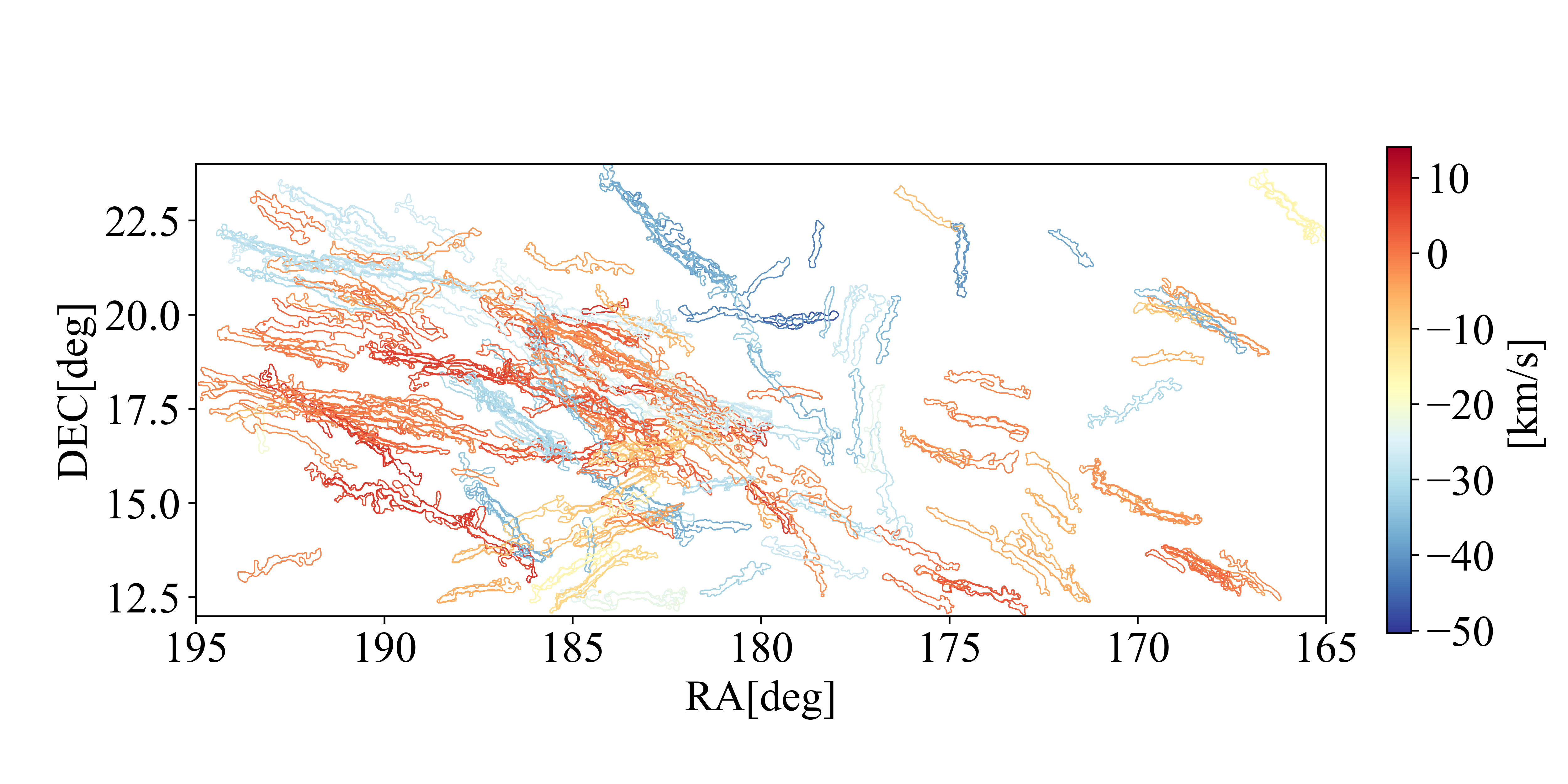}
\vspace*{-1cm}
\caption{The merged mask of all 269 extracted 3D filaments projected along the line-of-sight. Filaments shown have aspect ratios greater than 6:1. The color denotes the central channel of the detected velocity range for each filament.}
\label{fig:filaments}
\end{figure*}

\section{Data} \label{sec:data}
HI filaments are extracted from cubes of neutral hydrogen produced by the Galactic Arecibo L-Band Feed Array HI (GALFA-HI) Survey \citep{galfa}. GALFA-HI is a high angular and kinematic resolution survey of Galactic HI covering 13,000 deg$^2$ (approximately 1/3 of the sky) with 4$^\prime$ spatial resolution. We use the publicly-available GALFA-HI data which have a pixel size of 1 arcmin$^{2}$ and 0.184 km/s channel spacing over the velocity range $|v| < 188.4$~km/s. All velocities are measured in the LSR frame. The median rms noise is 352 mK at this resolution. 
In this paper, we focus on filaments residing in a high Galactic latitude region with an area of 360 deg$^2$ at $165^\circ<\alpha<195^\circ$ and $12^\circ<\delta<24^\circ$ which spans Galactic coordinates: l=[$210^\circ, 340^\circ$], b=[$59^\circ,90^\circ$]). 

In \Fig{fig:moment0}, we show an overlay of integrated intensity maps evaluated in two different velocity ranges over the spatial region we cover. The blue map is integrated over $-50 \leq v \leq -20$ $\rm{km\cdot s^{-1}}$, while the red is integrated over $-20 \leq v \leq 20$ $\rm{km\cdot s^{-1}}$. As shown, HI structures are visually distinct at different velocities.

We also employ the Planck 353~GHz (PR3.1) Stokes linear polarization maps \citep{planck353}. The native spatial resolution of the Planck data is FWHM=$4.9'$, comparable to GALFA-HI \citep{planck_xix}. For our analysis, we smooth the data to FWHM=$1^\circ$ to improve the signal-to-noise of the Planck data.

\section{Detecting 3D Filaments}  \label{sec:def}
We outline a procedure to extract 3D filaments from an emission cube. This algorithm is referred to as \texttt{fil3d} and will be described further and released to the public in a forthcoming work (Putman et al. in prep). To extract filamentary structures embedded in the diffuse ISM in the Milky Way, we first filter out large-scale diffuse Galactic emission. We apply an unsharp mask (USM), which effectively performs a high-pass spatial filter on the raw data. For this step, we first smooth each velocity slice of the data cube with a 30$^\prime$ Gaussian beam, then subtract the smoothed version from the original, and finally threshold the smoothed, subtracted data at zero. We then run \texttt{FilFinder} \citep{filfinder} to identify filamentary structures on each velocity channel of USM GALFA-HI data.

\texttt{FilFinder} employs the techniques of mathematical morphology to identify and segment two-dimensional filamentary structures over a wide dynamic range in brightness \citep{shih, filfinder}. To eliminate irregularities while maintaining a main structure, the algorithm first flattens and smooths the image, then applies an adaptive threshold to pick out all linear structures. These possible filament candidates are then trimmed to a pixel-wide skeleton with minimum connectivity using a Medial Axis Transform \citep{medial_axis}. The resulting skeletons are ``pruned" to be final filamentary structures by removing short branches that trace small deviations from the long axis of a filament. 
With the above procedures executed over the GALFA-HI velocity range $\lvert v \rvert \leq 188.4 \rm{km \cdot s^{-1}}$, we end up with collections of two-dimensional filamentary structures in individual velocity channels. We will refer to each \texttt{FilFinder}-detected 2D filament as a ``node". From these, we search for spatially overlapping node objects in neighboring velocity channels to obtain velocity-coherent structures. For instance, we start with a node at one channel, then search for objects at the next velocity channel that significantly overlaps (share 85$\%$ of pixels in common) with the first node. This search is executed one node at a time and continues until there are no objects found to yield a significant overlap in subsequent channels. 
As \texttt{fil3d} parameters, we assume the distance to be 100 parsecs and set the characteristic scale width as 0.1 parsec, which is the resolution of the GALFA-HI data at this distance. 

Each collection of 2D nodes forms a 3D filament and if a node is not matched with another in either adjacent velocity channel, it is rejected. We verify that all 3D filaments are unique in that each occupies a distinct set of spectral and spatial coordinates. The final projected shape of the 3D filament is defined by the line-of-sight sum of its constituent 2D nodes: we refer to this shape as the ``merged mask" of the 3D filament. 

\begin{figure*}
\centering
\includegraphics[width=\linewidth]{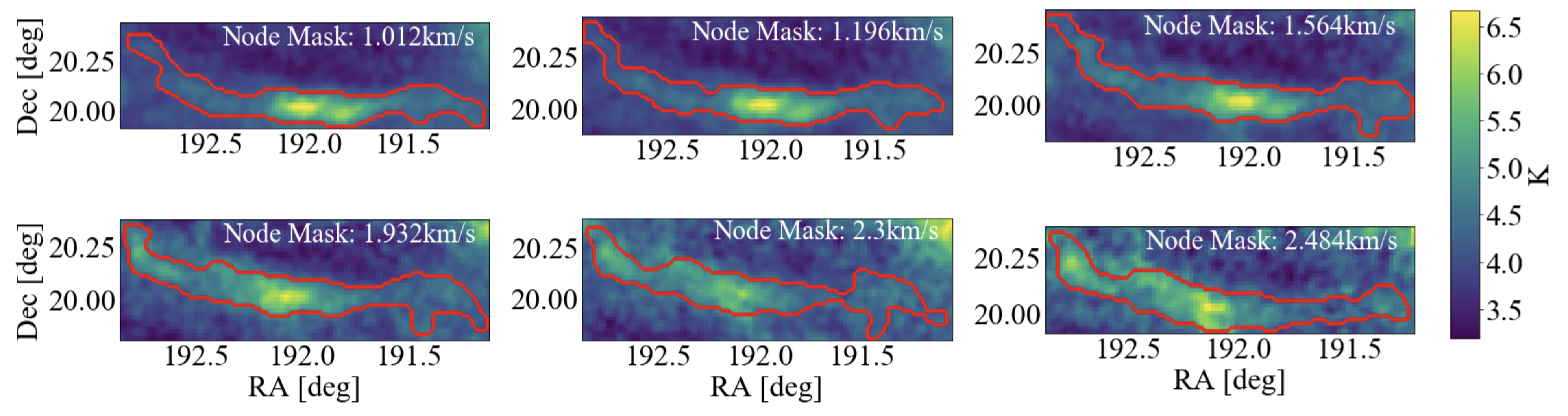}
\caption{Channel maps of one of the 3D filaments. The background image shows the brightness temperature from the raw data (before the USM is applied). The red contours outline the shapes of the nodes at velocity channels within the \texttt{fil3d} velocity width. This same filament will be used for several of the following plots.} 
\label{fig:channelmap}
\end{figure*}

In the high Galactic latitude region studied here, we initially find 325 3D filament candidates and apply the below filters to obtain 269 3D filaments for our final sample. For the final sample to ensure filament-like morphology, we only accept 3D structures with merged masks of aspect ratio (the ratio between the length of the major axis to its minor axis) of at least 6:1. We also apply a linewidth filter (see \S\ref{sec:vwidth}), which removes an additional 10$\%$ of the candidate filaments in our region. Roughly 1$\%$ of the HI flux in this region of the sky corresponds to 3D filaments.

Figure \ref{fig:filaments} shows the merged masks of our final selection of 3D filaments.  An example of the individual nodes (or channel maps) for a filament as found by \texttt{fil3d} is shown in Fig.~\ref{fig:channelmap}.

\begin{figure}
\centering
\includegraphics[scale=0.45]{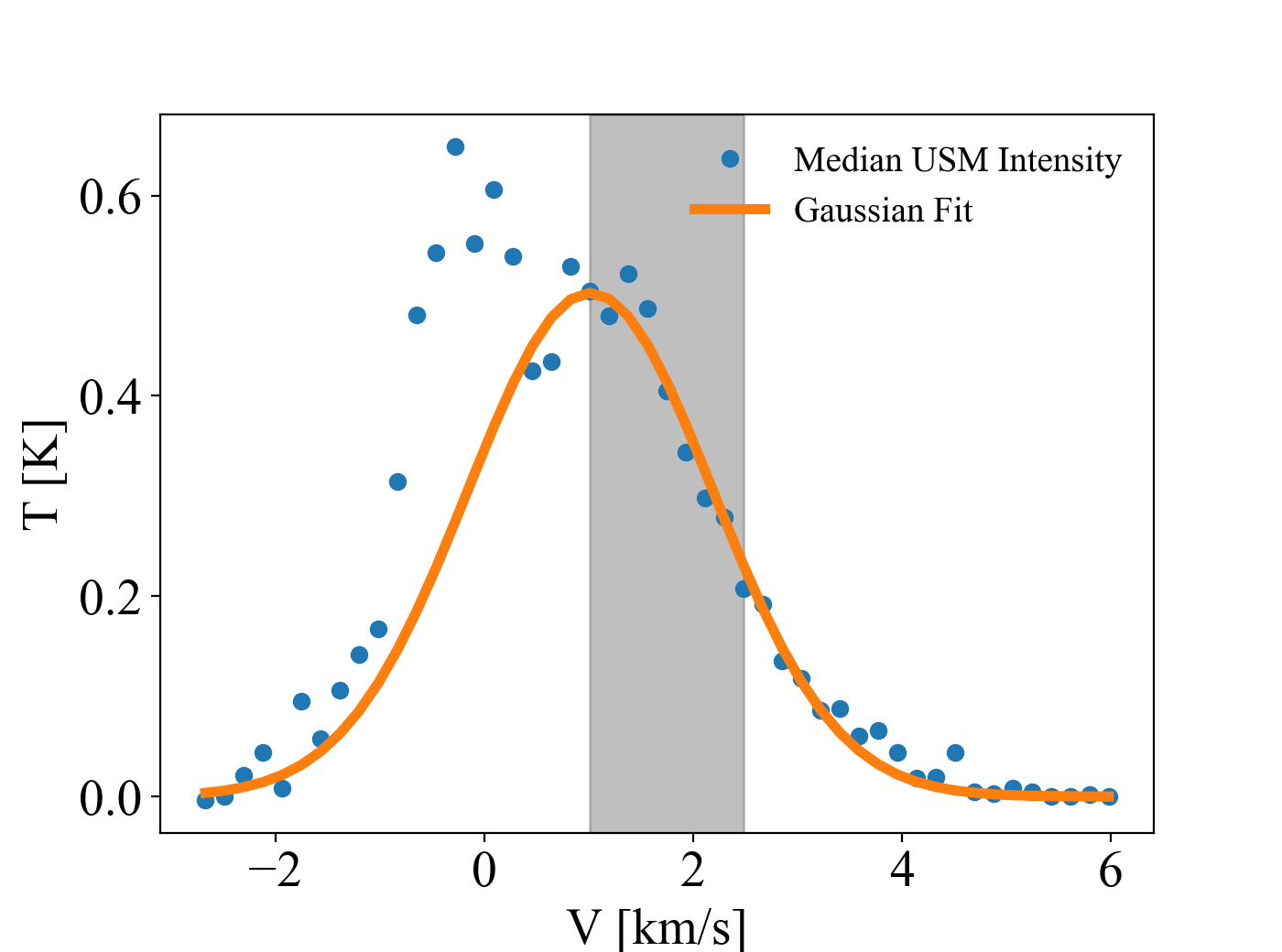}
\caption{A spectrum of the filament shown in \Fig{fig:channelmap}. The x-axis shows the velocity and the y-axis denotes the USM median intensity within a merged masked area of the filament (blue dots). The orange curve is the best-fit Gaussian of this velocity spectrum. The gray region indicates the \texttt{fil3d}-detected velocity range. The physical velocity width of a filament is defined as the full-width-half-max (FWHM) of the fitted Gaussian. The best-fit Gaussian peaks near the \texttt{fil3d}-detected velocity range, and we confirm via visual inspection that emission within the Gaussian-selected velocity range is associated with the \texttt{fil3d}-detected filament. }
\label{fig:vwidth_fit}
\end{figure}

\begin{figure*}
\centering
\includegraphics[width=\linewidth]{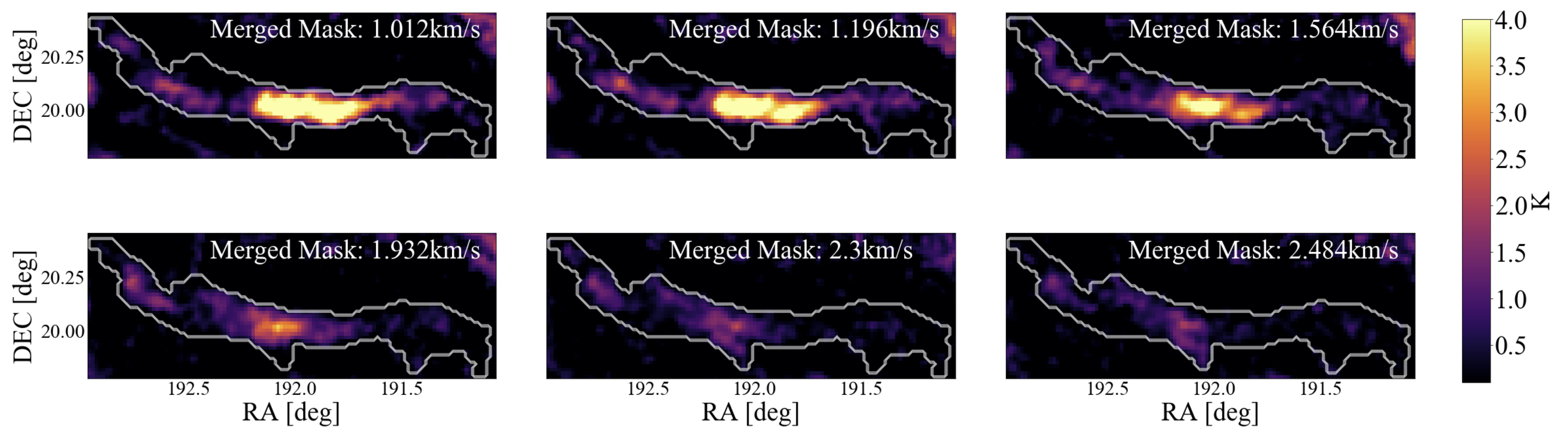}
\caption{Channel maps similar to \Fig{fig:channelmap}, but the white contours outline the merged mask area of the 3D filament (combined shape of all nodes) and the background is made with the USM data instead.}
\label{fig:merged_mask}
\end{figure*}

\begin{figure*}
\includegraphics[width=\linewidth]{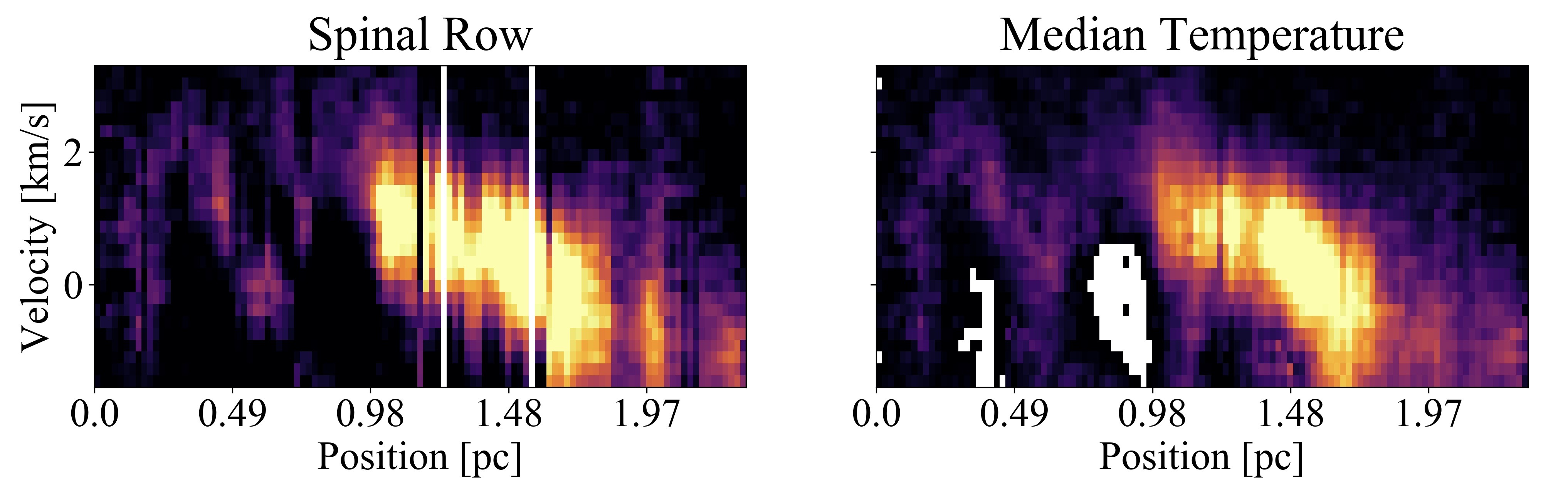}
\caption{Two position-velocity (PV) diagrams are built for each filament. The PV diagram on the left is based on the spinal pixels, while the right is built based on the median brightness temperature of each column for the merged mask (see text). The position (x-axis) denotes the projected length of the merged mask under an assumption that the distance is 100 pc. The velocity axis covers the FWHM range from \Fig{fig:vwidth_fit}. The color bar range is identical to \Fig{fig:merged_mask}.}
\label{fig:PV_example}
\end{figure*}

\section{Analysis Methods}\label{bigsec:methods}

\subsection{Linewidth Estimation}  \label{sec:vwidth}
The thermal linewidth is a useful indicator of the physical nature of ISM structures. \texttt{fil3d} catalogs filamentary structures which span greater than two velocity channels, and over half of the 3D filaments found occupy only two channels. 
Considering the fine spectral resolution of GALFA-HI (0.184 km/s), such a narrow velocity width deems unphysical. Cold gas in the Milky Way has a typical linewidth  of around 2-3~km/s \citep{kalberla_thermalwidth}. 

Upon inspection of the two-channel filaments, many were found to have additional emission within the merged mask in adjacent channels that was not captured by our criteria as clearly filamentary. Thus, rather than adopting the \texttt{fil3d} velocity range as the filament velocity width, we define a procedure for fitting a line profile that we consider to be more representative of all the emission associated with each filament.

\begin{figure*}
\centering
\includegraphics[width=0.9\linewidth]{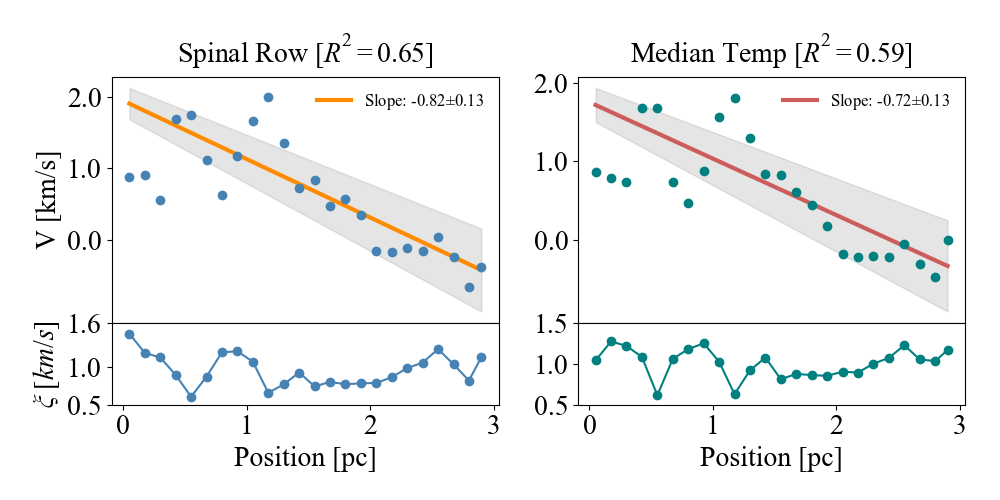}
\caption{The intensity-weighted mean velocity translated from the PV diagrams in \Fig{fig:PV_example} (top) and the square root of the second moment of velocity, $\xi$ (bottom), along the major axis of a filament. The x-axis denotes the projected filament length in parsecs (at an assumed distance of 100 pc) and the y-axis is the intensity-weighted mean velocity evaluated in each column of the PV diagram. The grey bands represent the model uncertainty derived from one sigma uncertainty of fitted parameters. The label indicates the best-fit slope magnitude (equivalent of gradient magnitude) and its uncertainty and has units of $\rm{km}\cdot\rm{s}^{-1}\cdot\rm{pc}^{-1}$. We evaluate $R^2$ to determine the goodness-of-fit and consider $R^2 > 0.5$ fits to have statistically significant gradients.}
\label{fig:slope_fit}
\end{figure*}

To derive the linewidth of each filament, we find the median USM intensity for each channel within the filament's merged mask area for channels $\pm 10$~km/s from the filament's central velocity as shown in Figure \ref{fig:vwidth_fit}. For this analysis, we re-bin the data into $4'$ pixels, to approximately match the GALFA-HI angular resolution. 
In Figure \ref{fig:vwidth_fit}, the blue dots denote the data points, and the orange line indicates a best-fit Gaussian to the data. The linewidth we adopt going forward for each filament is the full-width-half-max (FWHM) of the orange curve. Figure \ref{fig:vwidth_fit} shows two intensity peaks, one at $\approx$ 0 km s$^{-1}$ and another at $\approx$ 2 km s$^{-1}$. The peak near the \texttt{fil3d} detected range is associated with the selected filament, while the 0 km s$^{-1}$ peak corresponds to contamination from another filament.

We further implement a two-step examination to eliminate filaments from unassociated emission that is not coincident with the peak velocity in the intensity spectrum. As the first step, we check that the \texttt{fil3d}-detected channels are located within 1$\sigma$ of the peak of the fitted Gaussian. For filaments that meet this criterion, we then examine the individual channel maps to make sure that the filament emission is located within the merged mask area and does not extend significantly beyond it. Approximately ten percent of the total detected filaments are eliminated in these two-step examinations.

\subsection{Magnetic Field Orientation}  \label{sec:borientation}
The thermal emission from interstellar dust is linearly polarized because the short axes of dust grains are preferentially oriented parallel to the ambient magnetic field \citep{purcell_grain, Andersson_grain}. 
Thus, the linear polarization of this radiation is orthogonal to the plane-of-sky magnetic field orientation in the dusty ISM. To measure the magnetic field orientation towards HI filaments, we use Planck polarization maps \citep{planck353} at 353~GHz, a frequency dominated by thermal dust emission.  It is important to note that photometric dust polarization measures emission integrated over the line of sight and is therefore not  velocity-resolved as our 3D filaments are.

The polarization fraction (p) and polarization angle ($\psi$) in our analysis follow the IAU convention in Equatorial coordinates and are defined with the observed Stokes parameters (I,Q,U):
\begin{equation}
p = \frac{\sqrt{Q^2+U^2}}{I}, \\
\label{eq:pol-int}
\end{equation}
\begin{equation}
\psi = 0.5 \times \rm{arctan(-U,Q)}.
\label{eq:pol}
\end{equation}
To obtain the mean dust polarization fraction and polarization angle for an individual filament, we apply equations \eqref{eq:pol-int} and \eqref{eq:pol} respectively.  We define the plane-of-sky magnetic field orientation ($\phi$) to be rotated 90 degrees from the measured polarization angle ($\psi$).  We compute the magnetic field orientation for each filament by measuring the mean Stokes parameters within the merged mask area. 

The propagated statistical uncertainties (used in \S~\ref{sec:alignment}) are computed from the noise covariance matrices  and quantified in \cite{planck_xix} as 

\begin{equation}
\sigma_{\phi} = 28.65^{\circ}\sigma_P \times \frac{1}{P} \sqrt{\frac{Q^2\textbf{C}_{UU} + U^2\textbf{C}_{QQ} - 2QU\textbf{C}_{QU}}{Q^2\textbf{C}_{QQ}+U^2\textbf{C}_{UU} + 2QU\textbf{C}_{QU}}},
\end{equation}
where $\mathbf{C}_{QQ}$, $\mathbf{C}_{UU}$ are the internal variances and $\mathbf{C}_{QU}$ denotes the off-diagonal terms of the noise covariance matrix.  
The uncertainty on the polarized intensity is given by 
\begin{equation}
\label{eq:phi_uncertain}
\sigma_P^{2} = \frac{1}{P^2}(Q^2\textbf{C}_{QQ} + U^2\textbf{C}_{UU} + 2QU\textbf{C}_{QU}).
\end{equation}

\subsection{Filament Position-Velocity Diagrams} \label{sec:gradients}

Given the relative orientation of the filaments with respect to the magnetic field and the resolution of the HI observations, we focus on analyzing the velocity gradients along the long axis of the filaments. To compute the velocity gradient, we employ the position-velocity (PV) diagram of each filament. A PV diagram measures the distribution of velocities along a projected position and has been employed in understanding the large-scale kinematics of gaseous structures \citep{pv_rot, pv_exmp}. We construct PV diagrams here to determine the direction and magnitude of any velocity gradient. For consistency, we define the direction of the gradient to point from the lowest velocity to the highest velocity. 

To ensure we capture the internal kinematics of filaments, we construct two distinct PV diagrams with the USM data. The first PV diagram evaluates the intensity within spinal pixels $\boldsymbol{\tau}(x',y')$, or pixels that pass through the long axis spine (central region) of a 3D filament. \Fig{fig:merged_mask} shows the contour of the mask of a full 3D filament over the brightness temperature map of the filament at different velocity channels. 

For the first PV diagram, we construct the spinal pixels of 3D filaments by rotating the masks of the 3D filament to project onto a global horizontal axis. Here, the amount of rotation ($\theta$) is the angle between the spinal pixels of the non-rotated filament $\boldsymbol{\tau}(x,y)$ and the global horizontal axis. After the rotation, all pixels are projected onto grids of equal size. The spinal pixels are the central pixels in every column of this projected mask:



\begin{equation}
\boldsymbol{\tau}(x',y') = Median(\mathbf{R}(\theta)\cdot \boldsymbol{\tau}(x,y))
\end{equation}
where prime indicates the rotated frame and $\mathbf{R}(\theta)$ represents a rotation matrix written as 
\begin{equation}
\mathbf{R}(\theta) =    
 \left[
  \begin{array}{ c c }
     {\rm cos} \theta & {\rm -sin}\theta \\
     {\rm sin} \theta & {\rm cos}\theta
  \end{array} \right].
\end{equation}

In the second PV diagram, we evaluate the median brightness temperature along the long axis of a filament. As shown, a majority of the 3D filaments have irregular shapes that the spinal pixels may not always capture the full extent of a filament. Addressing this account, we evaluate the range of velocities at the median brightness temperature along a given column of the rotated merged mask.
In the right panel of Figure~\ref{fig:PV_example}, we show the range of velocities at the median brightness temperature along a given column of the rotated merged mask. 
As shown in \Fig{fig:PV_example}, not surprisingly, both the ``spinal row" and ``median brightness temperature" PV diagrams are highly correlated. This similarity shows that either method is broadly representative of the filament velocity structure.

\begin{figure}
\centering
\includegraphics[scale=0.37]{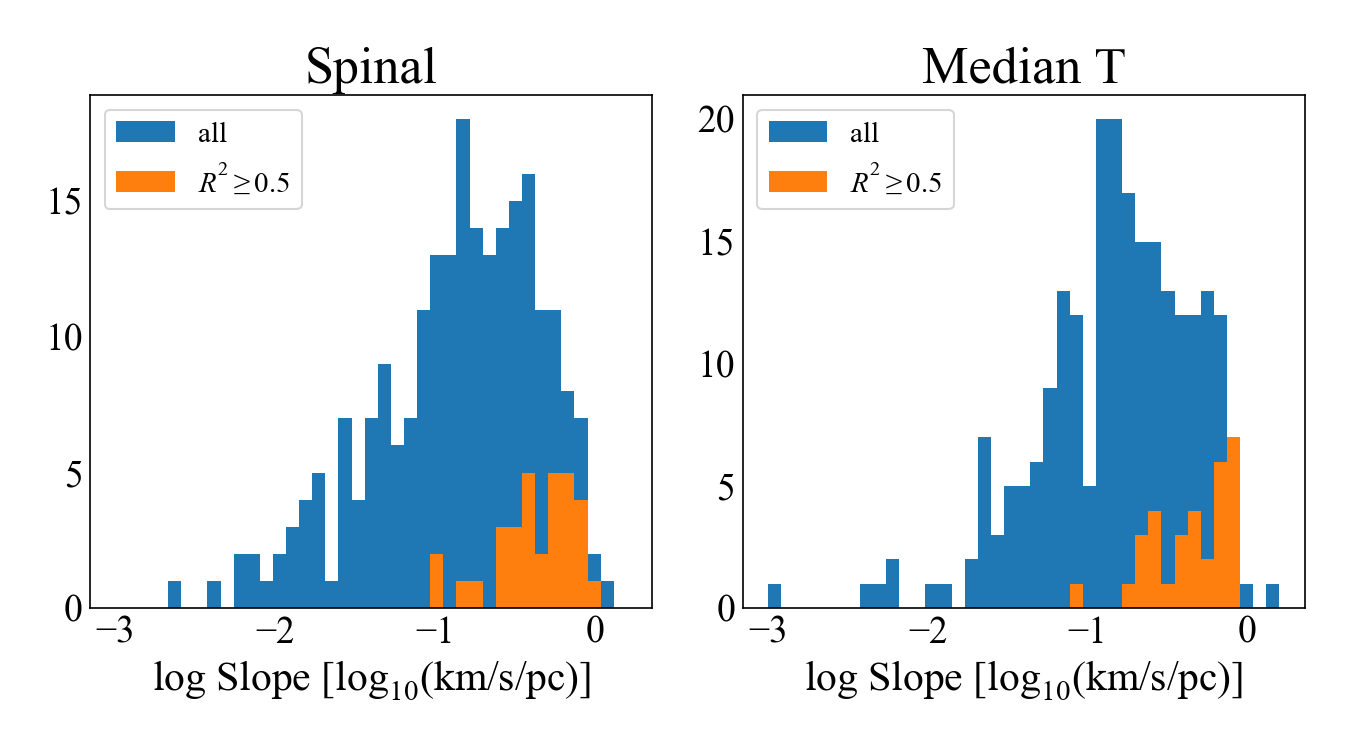}
\caption{Histograms of the velocity gradient slope magnitudes evaluated from spinal and median brightness temperature PV diagrams using a uniform distance of 100 pc. The high $R^2$ samples (orange) have preferentially higher magnitudes, around 0.1 to 1 km s$^{-1}$pc$^{-1}$, compared with the total distribution (blue).}
\label{fig:slope_magnitude}
\end{figure}

\subsection{Measuring Velocity Gradients} \label{grad_measurement}
We determine the presence (or lack) of velocity gradient along the major axis of filaments using PV diagrams from \S\ref{sec:gradients}. To extract a single velocity value along a filament and evaluate the magnitude and direction, we use the intensity-weighted mean velocity along each projected position (the position used as the x-axis of Figure~\ref{fig:PV_example}) and fit a slope to the points (as shown in \Fig{fig:slope_fit}) .
For example, the intensity-weighted mean velocity at position $P = p_j$ is expressed as
\begin{equation}
\boldsymbol{\omega}(p_j) =  \frac{\sum_{v_{\rm{fwhm}}} v \cdot \mathbf{I}(p_j,v)}{(\sum_{v_{\rm{fwhm}}}{\mathbf{I}(p_j,v)})},
\label{eq:iweigh_v}
\end{equation}
where $v_{\rm{fwhm}}$ above is evaluated over the FWHM from the velocity width fit in \S\ref{sec:vwidth} at a given position. 

In the top row of \Fig{fig:slope_fit}, we show the intensity-weighted mean velocity along the projected long axe of a filament (x-axis). With these data for each filament, we perform a weighted least squares regression to obtain a best-fit linear model (solid line). The second moment, $\boldsymbol{\xi}^2$, is used as the weight in the regression. At each pixel ($p_j$):
\begin{equation}
\boldsymbol{\xi} (p_j) = \sqrt{\frac{\sum_{v_{\rm{fwhm}}}(v-\boldsymbol{\omega}(p_j))^2 \mathbf{I}(p_j,v)}{\sum_{v_{\rm{fwhm}}} \mathbf{I}(p_j,v)}},
\label{mom1equat}
\end{equation}
which shares the same notation as Equation \ref{eq:iweigh_v}. We show the square root of the second moment ($\xi$) in the bottom row of \Fig{fig:slope_fit}.
The magnitude of the slope represents the magnitude of a potential velocity gradient, and the sign of the slope denotes the directional component parallel to the filament's long axis.


Not all filaments demonstrate velocity gradients (examples are shown in \Fig{fig:appendix_slopefit}). To select filaments that have significant velocity gradients, we evaluate the goodness of the linear fit to the data. This can be assessed from the coefficient of determination ($R^2$), which is the ratio of the variance explained by a fitted model to the total variance. The values for $R^2$ range from 0 to 1. An ideal model which perfectly explains all the variance in an observation will result in $R^2 = 1$. A poor fit, on the other hand, will result in a $R^2$ close to 0. We evaluate $R^2$ independently for the two PV diagrams described in \S\ref{sec:gradients} as both capture related yet separate kinematics of a given filament. We consider a filament velocity gradient significant if one of the PV diagrams has the $R^2$ slope fit greater than 0.5 (see Figure~\ref{fig:slope_magnitude}). 

\section{Results} \label{sec:results}
\subsection{Local and IVC Filament Populations} \label{sec:pop}
\begin{figure}
\includegraphics[scale=0.6]{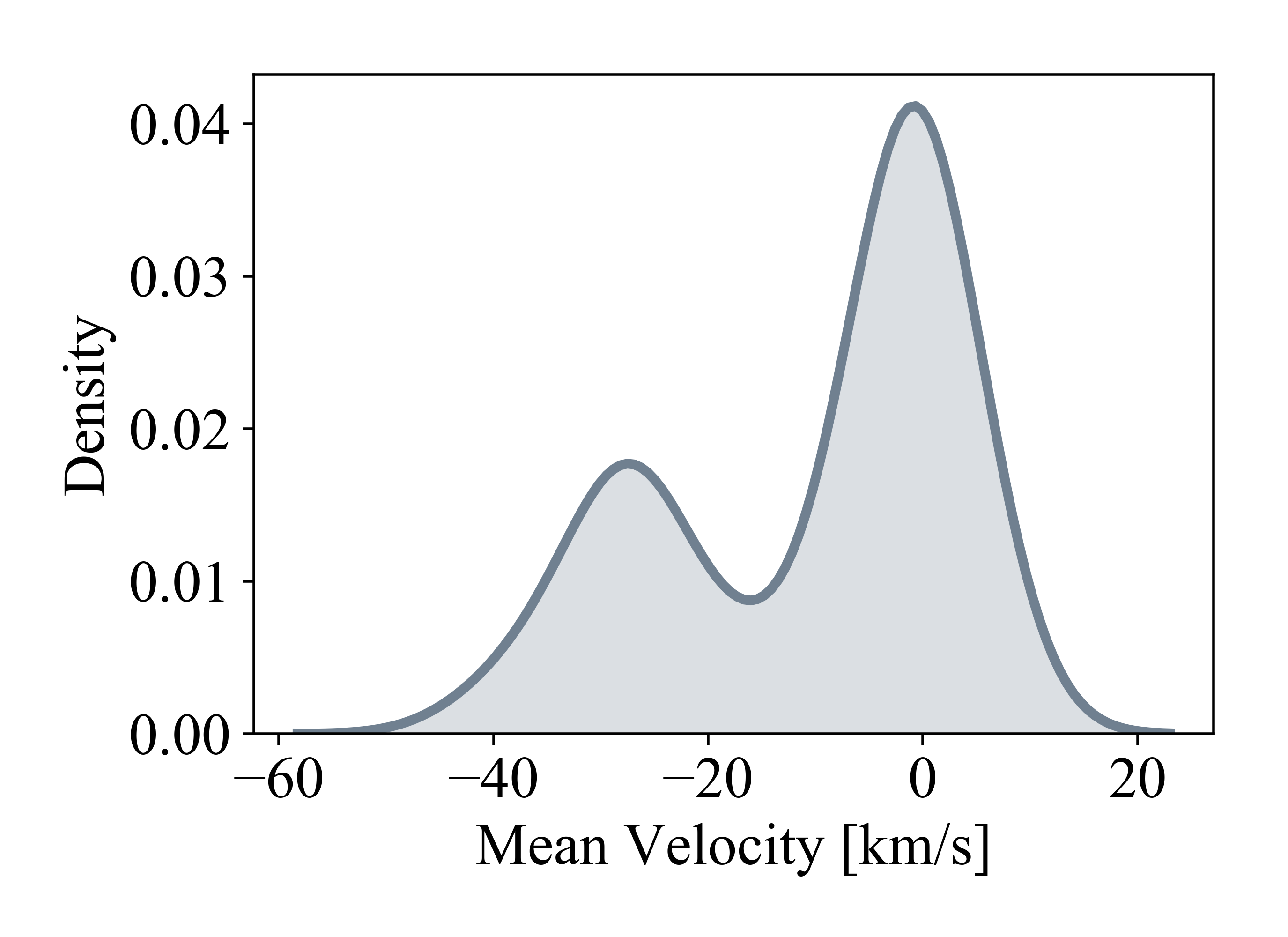}
\caption{A kernel density estimate (KDE) plot of central velocities of all HI filaments. The distribution shows a clear bi-modality, separating the local and IVC filaments in our sample.}
\label{fig:bimodal}
\end{figure}

\Fig{fig:bimodal} shows the central velocity distribution of the 3D filaments and suggests they can be separated into two distinct groups.
The first group, 157 filaments, has a central velocity near $v \approx 0$~km/s, which suggests the filaments belong to local gas co-rotating with the local standard of rest (LSR). The second group has 48 filaments with velocities $v<-30$~km/s.  This second group of filaments is likely part of a previously identifiedintermediate velocity cloud (IVC) that has velocities that deviate from a simple model of Galactic rotation \citep{wakker_ivcmap, putman_galactic_halo}. 

\begin{figure*}
\centering
\includegraphics[scale=0.48]{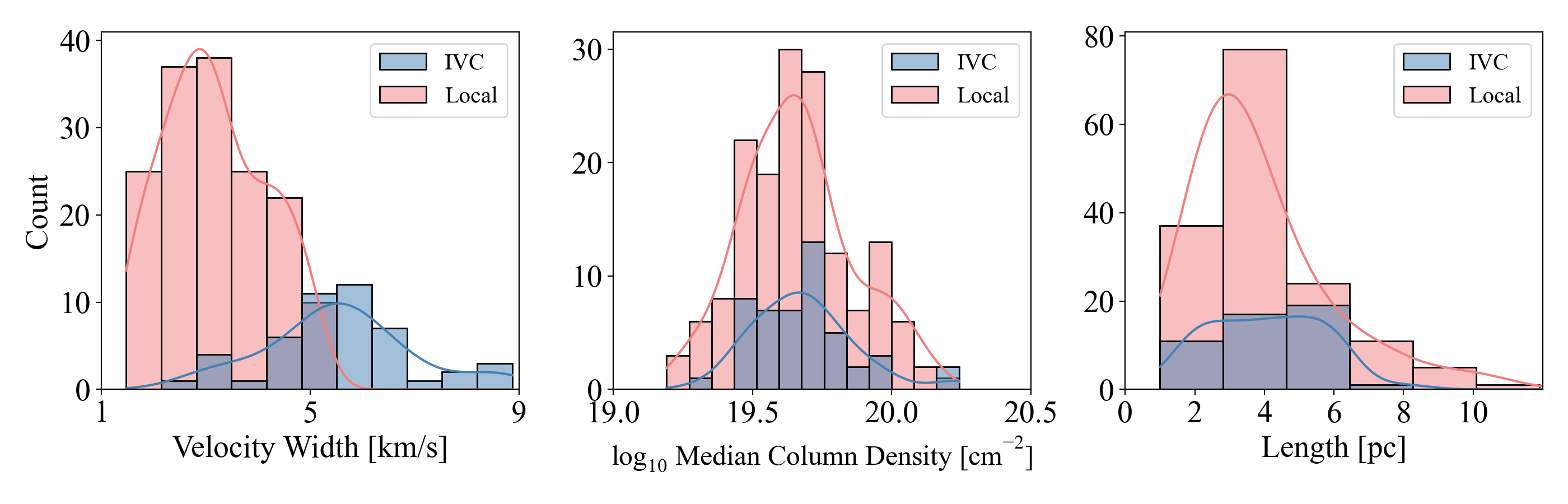}
\caption{\textit{Left}: Comparison of velocity widths between the local and IVC filaments derived from USM data. \textit{Center}: The median column density comparison. The column density is computed from the raw data using Equation \ref{eq:col_den} with the integral over the velocity width from the FWHM estimation.  \textit{Right}: The length of the filament major axes, when a uniform "fiducial" distance of 100 pc is assumed.}
\label{fig:biprops}
\end{figure*}

We assess the properties of the local and IVC filaments separately. In \Fig{fig:biprops}, we compare the line widths, median column densities, and lengths of the two populations. In general, we make use of the USM data because the raw data can be affected by some contamination from diffuse Galactic emission. The raw data is used when estimating column densities.

As shown on the left of \Fig{fig:biprops}, the IVC filaments tend to have larger velocity widths (linewidths of filaments are evaluated using the technique described in \S \ref{sec:vwidth}). The mean linewidth of the local filaments is 3.1~km/s, consistent with the typical thermal linewidths of cold HI structures in the Milky Way, and with the theoretical cold neutral medium (CNM) temperatures \citep{wolfire_1995, Kalberla_HI}.
The IVC filament linewidths peak at 6.2~km/s, indicating this is a population of warmer filaments. This would be expected for filaments directly associated with an IVC complex \citep{haud_ivcfit}.

We evaluate the column densities of the filaments and compare them in \Fig{fig:biprops}. The column density $N_{\rm{HI}}$ is computed with
\begin{equation}
N_{\rm{HI}} = 1.824 \cdot 10^{18} \int_{v_0}^{v_n} T_{b}(v) dv \  {\rm cm}^{-2},
\label{eq:col_den}
\end{equation}
where $T_{b}(v)$ is the brightness temperature for a given point within the merged mask area at one velocity channel $v$ and $\int |\ldots|$ indicates the moment 0 evaluated over the linewidth of each filament. As demonstrated from the center plot of \Fig{fig:biprops}, the median column densities of the two populations are comparable with a median of $\approx 10^{19.6}~\rm{cm^{-2}}$. These values are derived from the raw data because the USM data over-subtracted the column density. Thus the estimated column densities with the raw data may be biased high by the inclusion of diffuse Galactic emission un-associated with the filaments.

The rightmost panel of \Fig{fig:biprops} shows the major-axis filament lengths for each filament population, assuming all filaments are at a fiducial distance of 100 pc. The distributions of the two populations are fairly consistent. However, if the likely distance difference between the IVC and the local filaments is taken into account, the length distribution of the IVC filaments will shift further right.

\subsection{Alignment with the Magnetic Field}
The magnetic field orientation ($\phi$) of individual filaments is measured using the Planck 353~GHz Stokes parameter maps (see \S \ref{sec:borientation}). The left panel of \Fig{fig:ang_diff} demonstrates the difference between the mean magnetic field orientations ($\phi$) and spatial orientations ($\theta$) of filaments, while the right panel shows the measurement uncertainty evaluated from the smoothed Planck Stokes covariance using Equation \ref{eq:phi_uncertain} over the area of the filament merged mask.

\begin{figure}
\hspace{-0.8cm}
\centering
\includegraphics[scale=0.55]{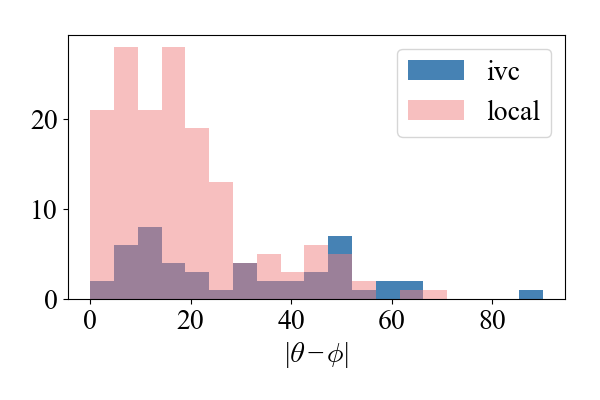}
\caption{The comparison in absolute difference distribution between the spatial orientations of the filaments ($\theta$) and the mean magnetic field orientations ($\phi$) from the Planck 353GHz data. Local filaments tend to be more aligned with the ambient magnetic field compared with the IVC filaments.}
\label{fig:bimodal_angle}
\end{figure}

Consistent with \cite{clark_rht}, \Fig{fig:ang_diff} demonstrates that the majority of the 3D HI filaments are well-aligned with the ambient magnetic field with relatively low $\phi$ uncertainties in general. 
However, we find different behavior when we compare the local and IVC populations. \Fig{fig:bimodal_angle} compares the relative orientations of the filaments and the magnetic field ($|\theta - \phi|$) for the two populations. As shown, the local filaments are well-aligned with the ambient magnetic field, and most perpendicular filaments in \Fig{fig:ang_diff} are identified to be the IVC filaments. The dust polarization traces the plane-of-sky component of the magnetic field orientation in a density-weighted integral along the line of sight.  

To further evaluate the relative orientation between the major axes of filaments and the magnetic field, we compute the extended projected Rayleigh statistics (PRS, \citealt{jow_prs}). The global PRS value ($Z_x$) quantifies the level of agreement between the orientations of the filaments and their average relative orientation with respect to the local magnetic field:
\begin{equation}
Z_x = \frac{\sum_{i}^{N} \rm{cos}|(\theta_i - \phi_i')|}{\sqrt{n/2}},
\end{equation}
where $\theta$ denotes a projection angle, $\phi'$ and $n$ are the mean magnetic field orientation measured in the same frame as $\theta$ and the number of filaments, respectively, and $Z_x >>0$ indicates strong parallel alignment while $Z_x << 0$ indicates strong perpendicular alignment. 

We obtain $Z_x=14.4$ and $\sigma_{Z_x}=0.94$ for the overall filament population. The number density averaged $Z_x$ of the local filaments is greater than that of the IVC filaments, which re-confirms what we visually inspect from \Fig{fig:bimodal_angle}: While the HI filaments at high Galactic latitudes are generally well-aligned with the ambient magnetic field, the local filaments are better aligned with the field. We find no significant correlation between the column densities of filaments and the level of magnetic field alignment, consistent with the findings for low-density dust filaments in \cite{alignment_planckxxxii}. 

\subsection{Filament Kinematics} \label{subsec:result-flow}
We examine the internal kinematics of the filaments by analyzing the velocity gradients. As mentioned in Section \ref{sec:gradients}, we extract velocity gradients parallel to major axes of filaments with two types of PV diagrams, then fit linear models to estimate the magnitudes and the directions of gradients.
To select filaments with significant velocity gradients, we compare the R-squared metric on the fitted linear models. Approximately 15 percent of the HI filaments have $R^2 > 0.5$ for at least one of two PV diagrams, and we consider those samples to demonstrate statistically significant velocity gradients. The estimated gradients from the two PV diagrams are highly correlated: the  Spearman correlation coefficient between the two methods is 0.96. The strong agreement between these two methods builds confidence that our gradient inference is robust to particular choices in the PV diagram construction.

A majority of filaments do not show significant velocity gradients along their length. One driving factor for this seems oscillatory patterns observed in the PV diagrams. 
Oscillations in the intensity-weighted mean velocity along the length of filaments, as seen in \Fig{fig:slope_fit} and multiple examples in the Appendix, can suppress the magnitude of a slope and yield a lower $R^2$ value.

Gradient magnitudes are computed assuming a fiducial distance of 100 pc, roughly the distance to the Local Bubble (LB) wall \citep{Lallement2022}.  Our velocity gradient measurements are limited by the GALFA-HI sensitivity and resolution: we are only sensitive above $10^{-2}$~km $ \rm{s}^{-1} \rm{pc}^{-1}$ for a typical filament. The magnitude distribution shown in \Fig{fig:slope_magnitude} demonstrates that the statistically significant gradients ($R^2 \geq 0.5$) all have a magnitude greater than the GALFA-HI resolution limit and have a median velocity gradient of 0.5 km$ \rm{s}^{-1}\cdot \rm{pc}^{-1}$. If the distance to the IVC filaments are taken account, their gradient magnitude would be smaller than $\mathcal{O}(0.1)$ km$\cdot \rm{s}^{-1}\cdot \rm{pc}^{-1}$.

\begin{figure*}
\vspace*{-0.5cm}  
\centering
\includegraphics[width=\linewidth]{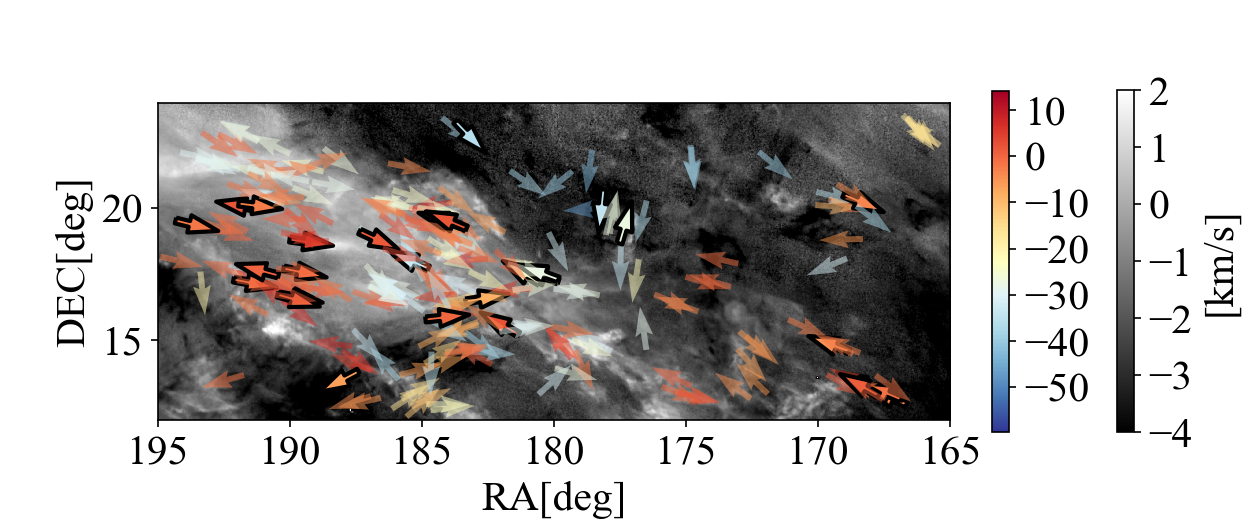}
\caption{Directional component of the filament velocity gradients expressed in the form of arrows which point to the direction of the higher velocity. The colors of the arrows indicate the filaments' central detected velocity. The darker outlined arrows denote the ones with significant velocity gradients with $R^2>0.5$. The background shows the first moment of the region evaluated from -10 to 10 km/s, but the color-scale is saturated for a better demonstration.} 
\label{fig:grad_vect}
\end{figure*}

\Fig{fig:grad_vect} illustrates the direction of gradients. The background is the moment 1 map evaluated in the [-10, 10] km/s range, and the arrows are positioned at each filament's location and point in the direction of increasing velocity. The colors of the arrows denote the central channel detected by \texttt{fil3d}, and the bold arrows show the filaments with $R^2 > 0.5$. Although no global trend is evident, some regions of local bulk flow seem to be captured (for example, the bottom right) by our analysis.   

The filaments with significant velocity gradients do not appear to correlate with other physical properties such as the filament's central velocity, column density, or magnetic field alignment; however, there is a weak correlation with filament length. Shorter filaments are more likely to have a statistically significant gradient, and the local filaments tend to be somewhat shorter than the IVC filaments on the sky. For instance, 92$\%$ of the local filaments with velocity gradients have their major axes shorter than approximately 5 parsecs (72 $\%$ of all local filaments have lengths less than 5 parsecs).

We attempted to evaluate the gradient perpendicular to the major axis of filaments; however, at most only two resolution elements are available across the minor axis of a filament. We performed a preliminary investigation of the kinematic structure perpendicular to the major axis over a region slightly wider than the filament masks.  Though a few filaments show gradients, we did not find clear evidence for perpendicular velocity gradients. To investigate a perpendicular gradient, much higher spatial resolution data of the filaments are needed.

\section{Discussion} \label{sec:discuss}
\subsection{Origin and Galactic Environment} \label{sec:gal_env}
The HI filaments can be placed into two groups based on their mean detected velocities as shown in \Fig{fig:bimodal}. The first group clusters around $v = 0$~km/s, which is consistent with gas associated with the solar neighborhood (we refer to this filament population as ``local"). The second group of filaments clusters around at $-60\rm{km}\rm{s}^{-1} < v <-30~ \rm{km}\rm{s}^{-1}$, and can be associated with an intermediate velocity cloud (referred to as ``IVC"). The bimodality in the filament population is robust to \texttt{fil3d} parameter choices.

\begin{figure*}
\centering
\includegraphics[width=\linewidth]{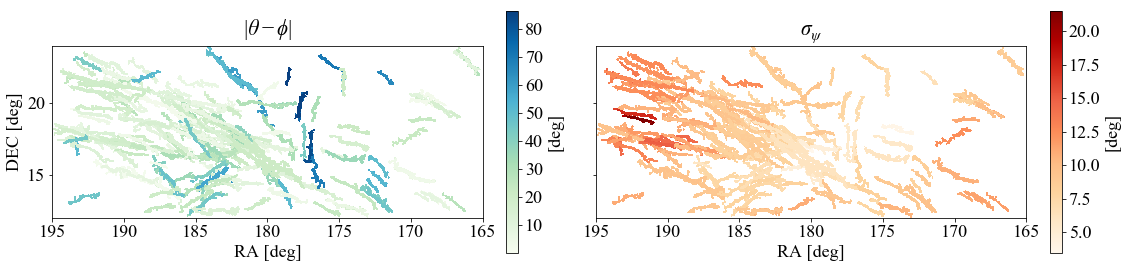}
\caption{Left: Difference between the filament orientation ($\theta$) and the magnetic field orientation inferred from the {\it Planck} 353\,GHz polarization observations. ($\phi$). Right: Mean angle uncertainty of the magnetic field orientation at the position of the filaments in the {\it Planck} data.}
\label{fig:ang_diff}
\end{figure*}

Given their low absolute velocities and position at high Galactic latitude, the ``local" filaments are likely located relatively nearby. A prominent feature of the nearby ISM is the wall of the LB. The LB is a low-density cavity of the ISM that surrounds the Sun \citep{bubble_map, Pelgrims_localbubble}. The column densities of our filaments imply they are located at a distance at least as far as the wall of the LB. The distance to the LB wall varies as a function of position in the sky, but is at least 100 pc away in most directions \citep[e.g.][]{cox_reynolds_lcdistance, snowden_localbubbleHI, murray2020_3dcloud, Lallement2022}: thus we set 100 pc as a lower-limit for filaments' estimated distance.

The formation of the local filaments is plausibly be linked to the formation of the LB. The winds from massive stars and explosions from nearby supernovae \citep{cox_sn_localbubble}, perhaps from the Sco-Cen association \citep{crutcher_ism}, injected the energy needed to stretch the cavity wall and redistribute the interstellar medium over spatial scales of a few hundred parsecs. In this process, filamentary structures can be created from the compressed interstellar magnetic fields in the walls or shells of HI gas shaped by the expanding bubbles \citep{alves_localbubble, local_buble_sn}. Under this scenario, the projected magnetic field follows the curvature of the expanding bubble, which leads to a large-scale correlation between HI geometry and magnetic field orientation. The existence of IVC filaments with similar properties, however, suggests that if the above bubble-linked formation mechanism is correct, it must be a sufficient, but not necessary, filament formation catalyst.

Our IVC filaments directly overlap with the large IVC complex called the Intermediate-Velocity Arch (IV Arch), which stretches from $\ell \approx 115^\circ$, $b \approx 35^\circ$ to $\ell \approx 200^\circ$, $b \approx 70^\circ$ \citep{wakker_ivcmap}. The IV Arch is at a z-height above the Galactic plane between 0.8 and 1.5 kpc \citep{kunz_danly_ivarch}, and has a local maximum N$_{\rm{HI}}$ column density in the spatial and kinematic region of our IVC filaments \citep{ivc_knude}. The physical environment of the IV Arch is not well-known, but its physical properties, including the magnetic field, local energy sources, and ISM composition, are unlikely to be identical to those of the LB. Despite these differences, 3D filaments are still found, indicating a variety of physical conditions can produce them.

 The difference in distance between the IVC and the LB leads to a natural size disparity between the two filament populations. The rightmost panel of \Fig{fig:biprops} shows that the two filament groups have a similar length, however, the IV Arch is approximately ten times further away than the neutral wall of the LB, which leads the IVC filaments to have lengths of around tens of parsecs. The linewidth difference between the two populations may represent the lower-pressure environment of the IVCs. Future work that maps filaments across the sky at a range of velocities will be key to further investigating the properties of HI filaments in different Galactic locations.  

\subsection{Implications of Filaments' Magnetic Field Alignments} \label{sec:alignment}

\Fig{fig:ang_diff} demonstrates that a majority of the HI filaments are well aligned with the plane-of-sky magnetic field orientation inferred from the Planck dust polarization angle. This agrees with previous analyses of HI filaments \citep{clark-pol, clark_hensley}, which find that structures in the diffuse medium are preferentially oriented to the local magnetic field. These previous works quantified the orientation of linear HI structures in individual velocity channel maps using the Rolling Hough Transform \citep{clark_rht}. Distinct from previous work, we explicitly measure the orientation of three-dimensional, velocity-coherent HI filaments. 

When comparing the relative magnetic field orientations between the local and IVC filament populations, we find the IVC filaments are less-aligned to the Planck 353 GHz magnetic field. This is perhaps not surprising.  
Polarized dust emission is a line-of-sight (LOS) integrated quantity. Along a single LOS, multiple layers of dust clouds with different spectral energy distributions (SEDs) and magnetic properties may exist, each contributing to the measured polarization angle \citep{Clark:2018, pelgrims_pol}.  While ``local" gas within the LB has a relatively uniform galactic dust-to-gas emission ratio with a moderate HI column density \citep{jones_local_dust}, the dust content significantly decreases for more distant gas clouds (e.g. IVCs and HVCs) \citep{reddening_localbubble}. A lower dust content may result from lower metallicities, usually related to the amount of dust, of distant clouds \citep{wakker_hvc}, or less heating from the interstellar radiation field due to their distance from the Galactic disc \citep{saul_dust_to_gas}. 

The density-weighted dust polarization at high Galactic latitudes is dominated by the local ISM, because the mean dust column of the local ISM is approximately twice that of the IVCs \citep{btomogrphy_panopoulou}. In other words, a lack of alignment between the dust polarization and the IVC filaments does not necessarily indicate the IVC filaments are not well-aligned with their \textit{local} magnetic field.
For instance, \cite{btomogrphy_panopoulou} estimated the plane-of-sky magnetic field orientation as a function of distance using stellar distance and starlight polarization measurements. They found that two clouds at different distances (IVC and LVC) exhibit significant differences in column density and polarization properties; not only do the two clouds differ in $p$, but also the mean magnetic field orientation at the distance of each cloud differs by 60$^\circ$. \citet{clark_hensley} found that the linear HI structures at the velocities coincident with these two clouds agree well with the magnetic field orientations measured by \citet{btomogrphy_panopoulou}. 
In order to examine the relative orientation of the IVC filaments to their ambient magnetic field, this type of tomographic investigation \citep[e.g.,][]{Pelgrims:2023} is required to disentangle the magnetic field structures along the LOS.

\subsection{Velocity Gradients} \label{sec:flow_bfield}
A number of theoretical models predict velocity gradients along various ISM filaments. HI filaments, if created by thermal instability and turbulent compression and shear, are aligned with the local magnetic field due to the turbulent shear strain induced at the shock front \citep{Inoue_Inutsuka, ntormousi_filform}. Some literature predicts bulk gas motions along filaments as the magnetic field directs the assembled flow along the field lines \citep{crutcher_Brho, Tritsis_striations}.

In our study, only approximately 15\% of the identified HI filaments demonstrate small velocity gradients along their major axes. Because the majority of filaments do not exhibit clear velocity gradients, we conclude that long-axis velocity gradients are not a ubiquitous characteristic of this filament population. Furthermore, the filaments' velocity gradients are not strongly correlated with their magnetic field alignments or column densities. Only a weak anti-correlation between the length of filament and velocity gradients is observed.

The viewing angle between the 3D filament orientation and the line of sight may contribute to an anti-correlation between filament length and velocity gradient magnitude \citep{fernandez_vgrad, chen_filgrad}. For fixed values of gas velocity gradient and filament length, filaments oriented more parallel to the line of sight should exhibit stronger radial velocity gradients and have shorter plane-of-sky extents. 
Although we see a hint of this expected anti-correlation in the data, we cannot conclude this trend is physically meaningful since most of our filaments do not exhibit clear velocity gradients in the first place. Moreover, we find that longer filaments are more likely to include knots of emission with more complicated velocity structures, as seen in Figure~\ref{fig:slope_fit} and \ref{fig:appendix_slopefit}, which results in a lower $R^2$ score. This physically or unphysically associated emission that is often present in the PV diagrams of longer filaments dilutes the magnitude of the velocity gradients and affects the goodness of fit.

Periodic velocity structures, somewhat similar to what is seen here, are also reported along the lengths of molecular filaments \citep{hacar_filform, Barnes2018, liu_oscillation, henshaw_vfluct, hacar_sf}. Velocity oscillations are present over different scales in the molecular filaments, and their origin is often assumed to be related to small-scale gravitational accretion or outflows of young stellar objects \citep{hacar_fragmentation, liu_oscillation, henshaw_vfluct}. We note that the mean gradient magnitudes between the molecular and our HI filaments are similar \citep{Goodman_vgrad, hacar_fragmentation, fernandez_vgrad, jimenez_vgrad, Dhabal_vgrad, chen_filgrad}. Although a strong correlation between the HI and $\rm{}^{13}$CO gas velocities are noted in molecular cloud candidates \citep{soler_molecular_grad}, our HI filaments are not self-gravitating, nor near star-forming regions. Any similarity between the kinematic structure of HI and molecular filaments must either be coincidental or related to other physics. Further studies are needed to explore the similarity between the velocity structures of atomic and molecular filaments.

\section{Conclusion} \label{sec:conclusion}
In this work, we study the kinematics and magnetic field alignment of 3D HI filaments at high Galactic latitude. The highlights of our findings can be summarized as follows.
\begin{itemize}

  \item We use a new filament-finding algorithm, \texttt{fil3d}, which searches for velocity-coherent filamentary structures. \texttt{fil3d} first finds filamentary (``node") objects in every velocity channel with \texttt{FilFinder} \citep{filfinder}, and then constructs three-dimensional filaments by extending nodes that significantly overlap with neighboring velocity channels. We run \texttt{fil3d} on GALFA-HI in a high Galactic latitude region, and identify 269 3D HI filaments after aspect ratio and velocity profile filtering.

  \item We observe our 3D filaments can be separated into two groups based on their mean detected velocities. The two groups differ in line widths, magnetic field alignments, and physical sizes, but share similar morphological properties. The results suggest the two groups of filaments originate from the Local Bubble and the IV-Arch respectively.
  
  \item We derive physical line widths of HI filaments by fitting a Gaussian to the USM intensity spectrum of individual filaments. The estimated velocity widths agree well with those of CNM structures. The typical linewidth of local and IVC filaments are 3.1km/s and 6.2km/s, respectively.
  
  \item We find the local HI filaments are well-aligned with the ambient magnetic field measured from the Planck 353GHz data. IVC filaments do not show the same level of alignment. This difference is likely due to the fact that the polarized dust emission is LOS integrated and the IVC filaments do not dominate the column.  A further tomographic effort is needed to disentangle the magnetic field structures along the line of sight.
  
  \item We develop a method of assessing filament velocity gradients from two types of PV diagrams. We find 15 percent of our filaments show significant velocity gradients along their long axes with their typical gradient amplitudes ranging between 0.1 to 1.2 $\rm{km}\cdot\rm{s}^{-1}\rm{pc}^{-1}$.
  
\end{itemize}

The results of this work show the importance of velocities and spectral and spatial resolution in studies of HI filaments. This paper presents the first finding of the alignment of velocity-coherent filamentary structures with the magnetic field and the presence of these structures at the Milky Way's disk-halo interface in the IVCs. Future tomography of the Galactic magnetic field will provide further insight into the alignment of filaments  with the magnetic field at varying distances \citep{pasiphae}. The finding that HI filaments do not typically display long-axis velocity gradients sets constraints on theoretical models of diffuse filament formation. 

\section*{Acknowledgements}
The authors thank helpful discussions with Blakesley Burkhart, Chang-Goo Kim, Mordecai-Mark Mac Low, Lorenzo Sironi, Snezana Stanimirovic, and Jacqueline van Gorkom. D.A.K thanks Eric Korpela for help with the GALFA-HI data cube. 
This work was partly supported by the National Science Foundation under Grant No. AST-2106607.

This project makes use of astropy \citep{astropy:2013, astropy:2018}, FilFinder\citep{filfinder}, healpy \citep{healpy1, healpy2}, numpy and scipy \citep{scipy}, matplotlib \citep{matplotlib}, and statsmodel \citep{statsmodel}.

\section*{Data availability}
This publication utilizes data from Galactic ALFA HI (GALFA-HI) survey data set obtained with the Arecibo L-band Feed Array (ALFA) on the Arecibo 305 m telescope (available at https://purcell.ssl.berkeley.edu/). This paper also makes use of observations obtained with Planck (http://www.esa.int/Planck), an ESA science mission with instruments and contributions directly funded by ESA Member States, NASA, and Canada. 
The data that support the plots within this paper and other findings of this study are available from the corresponding author upon requests.


\bibliographystyle{mnras}
\bibliography{hifil22}

\appendix

\section{Gallery of Velocity Gradients}
\begin{figure*}
\centering
\includegraphics[scale=0.65]{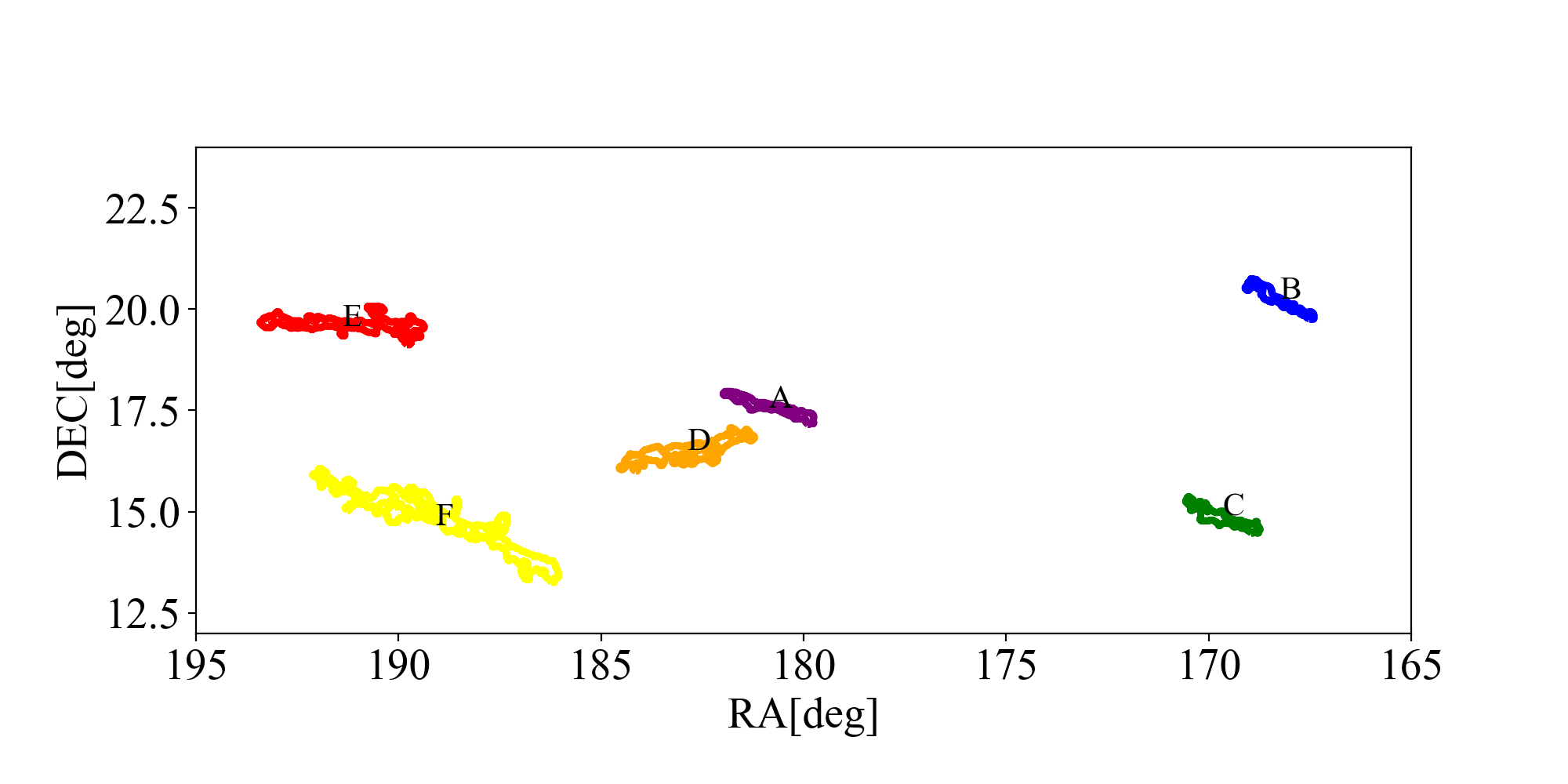}
\caption{Location of Reference Filaments}
\label{fig:appendix_map}
\end{figure*}

\begin{figure*}
\centering
\includegraphics[scale=0.42]{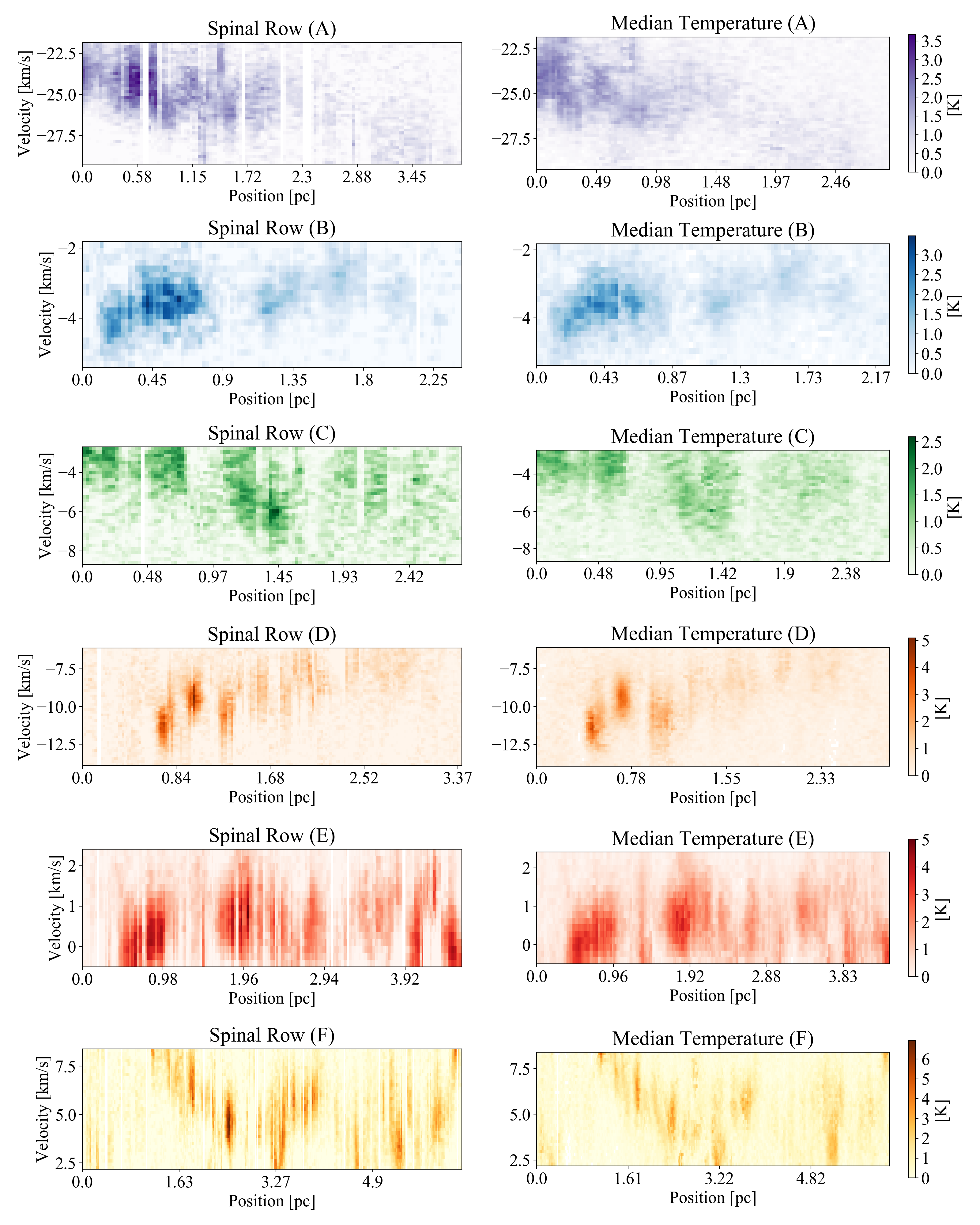}
\caption{PV diagrams of referenced filaments. We assume the distance to these filaments to be 100pc.}
\label{fig:appendix_pv}
\end{figure*}

\begin{figure*}
\centering
\includegraphics[scale=0.43]{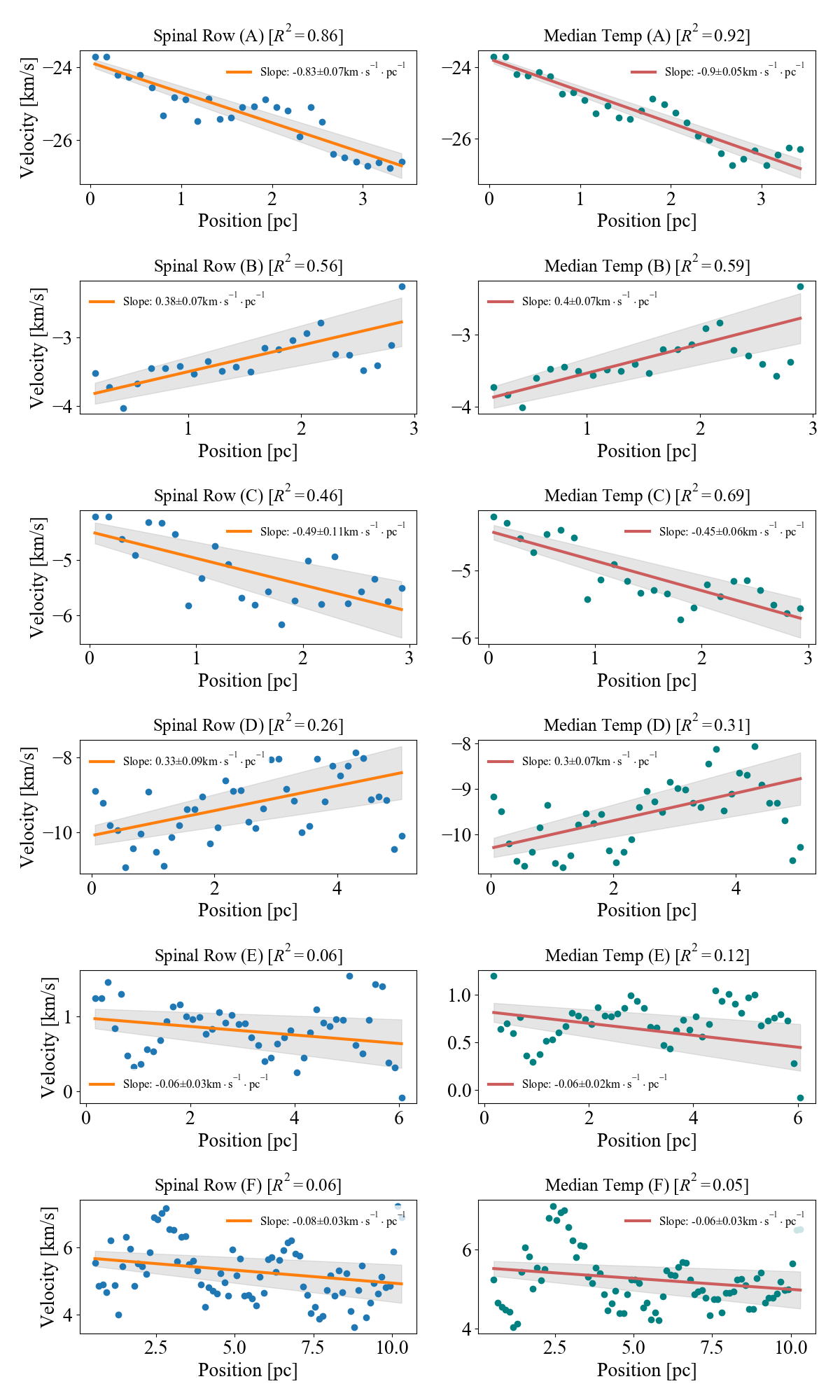}
\caption{Slope fits for the referenced filaments.}
\label{fig:appendix_slopefit}
\end{figure*}

Here we show PV diagrams of HI filaments located in different parts of the sky as shown in Fig.\ref{fig:appendix_map}. The letters on the filaments in Fig.\ref{fig:appendix_map} correspond to following PV diagrams. Fig.\ref{fig:appendix_pv} shows the PV diagrams using the two methods discussed in the text, and Fig.\ref{fig:appendix_slopefit} shows the intensity-weighted mean velocity translated from the two types of PV diagrams. The $R^2$ is evaluated for each method and the grey band represents the one sigma uncertainty of the fitted models (see text).

\bsp	
\label{lastpage}
\end{document}